\documentclass{revtex4}

\usepackage{graphicx}
\begin{document}
\bibliographystyle{revtex}


\hfill{OUNP-2001-05}

\hfill{December 2001}

\title{QCD Studies at the High-energy Linear $e^+e^-$ Collider}



\author{P.N. Burrows}
\email[]{p.burrows@physics.ox.ac.uk}
\affiliation{Oxford University}


\date{\today}

\begin{abstract}
I summarise the QCD programme at the high-energy $e^+e^-$
linear collider, as reported in the TESLA TDR and Linear Collider Physics Resource Book.
\end{abstract}

\maketitle

\section{Introduction}

Strong-interaction measurements at the Linear Collider (LC)
will form an important component of the physics 
programme. The collider 
offers the possibility of testing QCD at high energy scales
in the experimentally clean, theoretically tractable
$e^+e^-$ environment. In addition, virtual $\gamma\gamma$ interactions
will be delivered free by Nature, and a dedicated $\gamma\gamma$
collider is an additional option, allowing detailed measurements of
the relatively poorly understood photon structure.
The benchmark physics main topics are:

\begin{itemize}

\item
Precise determination of the strong coupling $\alpha_s$.

\item
Measurement of the $Q^2$ evolution of $\alpha_s$ and constraints on the GUT scale.

\item
Measurement of the total $\gamma\gamma$ cross section and 
the photon structure function; these issues are discussed elsewhere~\cite{albert}.

\end{itemize}

\section{Precise Determination of $\alpha_s$}

The current precision of individual $\alpha_s$
measurements is limited at best to several per cent~\cite{alphasrev}.
Since the uncertainty on $\alpha_s$ translates directly into an uncertainty 
on perturbative QCD (pQCD) predictions, especially for high-order multijet 
processes, it would be desirable to achieve much better precision. 
In addition, since the weak and electromagnetic couplings are known
with much greater relative precision, the error on $\alpha_s$ represents the
dominant uncertainty on our `prediction' of the scale for
grand unification of the strong, weak and electromagnetic forces~\cite{gut}.

Here we will refer to the conventional yardstick of
$\alpha_s$ quoted at the $Z^0$ mass scale,
$\alpha_s(M_Z)$, unless explicitly stated otherwise.
Several techniques for $\alpha_s(M_Z)$ determination will be available
at the LC:

\subsection{Event Shape Observables}

The determination of $\alpha_s(M_Z)$ from event `shape' observables that are
sensitive to the 3-jet nature of the particle flow has been
pursued for 2 decades and is generally well understood~\cite{philalp}. 
In this method one usually forms a differential distribution, 
makes corrections for detector and hadronisation effects,
and fits a pQCD prediction to the data, allowing $\alpha_s(M_Z)$ to vary.
Examples of such observables are the thrust, jet masses and jet rates.

The latest generation of such $\alpha_s(M_Z)$ measurements, from SLC and LEP, has
shown that statistical errors below the 1\% level can be obtained
with samples of a few tens of thousands of hadronic events.
With the current LC design luminosities of 
a few $\times$ $10^{34}$/cm$^2$/s at 500, 800 or 1000 GeV, 
hundreds of thousands of e$^+$e$^-$ $\rightarrow$ $q\overline{q}$ 
events would be produced each year,
and a statistical error on $\alpha_s(M_Z)$ below the 0.5\% level could
be achieved. 

Detector systematic errors, which relate mainly to
uncertainties on the corrections made for acceptance and resolution
effects and are observable-dependent, 
are under control in today's detectors at the $\Delta\alpha_s(M_Z)$ = 
1-4\% level~\cite{otmar}. 
If the LC detector is designed to
be very hermetic, with good tracking resolution and efficiency,
as well as good calorimetric
jet energy resolution, all of which are required for the search
for new physics processes, it seems reasonable to 
expect that the detector-related uncertainties can be beaten down to
the $\Delta\alpha_s(M_Z)$ $\simeq$ 1\% level or better.

e$^+$e$^-$ $\rightarrow$  $Z^0 Z^0$, $W^+W^-$, or $t\overline{t}$ events 
will present significant backgrounds to  $q\overline{q}$
 events for QCD studies, and the
selection of a highly pure $q\overline{q}$  event sample will not be quite
as straightforward as at the $Z^0$ resonance. The application of kinematic cuts 
would cause a significant bias to the event-shape
distributions, necessitating compensating corrections at the level of
25\%~\cite{bias}. 
More recent studies have shown~\cite{schumm} that the majority of
$W^+W^-$ events can
be excluded without bias by using only events produced with 
right-handed electron beams 
for the $\alpha_s(M_Z)$ analysis. Furthermore, the application of
highly-efficient $b$-jet tagging can be used to reduce the $t\overline{t}$ 
contamination to the 1\% level. After statistical subtraction of the
remaining backgrounds (the $Z^0 Z^0$ and $W^+W^-$ event properties
have been measured accurately at SLC and LEPI/II),
the residual bias on the event-shape distributions is expected to
be under control at the better than 1\% level on $\alpha_s(M_Z)$.

Additional corrections must be made for the effects of the smearing of the
particle momentum flow caused by hadronisation.
These are traditionally evaluated using Monte Carlo models.
The models have been well tuned at SLC and LEP and are widely 
used for evaluating systematic
effects. The size of the correction factor, and hence the uncertainty, 
is observable dependent, but the `best' observables measured at the
$Z^0$ have 
uncertainties as low as $\Delta\alpha_s(M_Z)$ $\simeq$ 1\%. Furthermore, one expects
the size of these hadronisation effects
to diminish with c.m. energy at least as fast as 1/$Q$.
Hence 10\%-level corrections at the $Z^0$ should dwindle to 
1\%-level corrections at $Q$ $\geq$ 500 GeV, and the
associated uncertainties will be substantially below the 1\% level
on $\alpha_s(M_Z)$. This has been confirmed by explicit simulations using 
PYTHIA~\cite{otmar}.

Currently pQCD calculations of event shapes 
are available complete only
up to $O(\alpha_s^2)$, although resummed calculations are available for some
observables~\cite{resum}. 
One must therefore 
estimate the possible bias inherent in measuring
$\alpha_s(M_Z)$ using the truncated QCD series.
Though not universally accepted, it is customary to estimate this from
the dependence of the fitted  value on the QCD renormalisation scale,
yielding a large and dominant uncertainty of about $\Delta\alpha_s(M_Z)$ $\simeq$
$\pm$6\%~\cite{philalp}.
Since the missing terms are $O(\alpha_s^3)$, and since 
$\alpha_s$(500 GeV) is expected to be about 25\% smaller than $\alpha_s(M_Z)$, 
one expects the uncalculated contributions to be
almost a factor of two smaller at the higher energy. However,
translating to the yardstick $\alpha_s(M_Z)$ yields an uncertainty of
$\pm$5\%, only slightly smaller than currently.
Therefore, although a 1\%-level  measurement is possible experimentally,
it will not be realised unless $O(\alpha_s^3)$ contributions are
calculated. There is reasonable expectation that this will be achieved within
the next 5 years~\cite{zvi}.
 
\subsection{The $t\overline{t}(g)$ System}

The dependence of the e$^+$e$^-$ $\rightarrow$ $t\overline{t}$
production cross section, $\sigma_{t\overline{t}}$, on the top-quark mass, $m_t$,
and on $\alpha_s(M_Z)$ are discussed elsewhere~\cite{tdr}. 
In order to optimise the precision on the $m_t$ measurement near threshold
it is desirable to input a precise $\alpha_s(M_Z)$ measurement from elsewhere.
Furthermore, 
the current theoretical uncertainty on $\sigma_{t\overline{t}}$ translates into
$\Delta\alpha_s(M_Z)$ = $\pm10\%$. Hence, although extraction of $\alpha_s(M_Z)$ from 
$\sigma_{t\overline{t}}$ near threshold may provide
a useful `sanity check' of QCD in the $t\overline{t}$  system,
it does not appear currently to offer the prospect of a competitive 
measurement.
A preliminary study has also been made~\cite{werner} of the determination
of $\alpha_s(M_Z)$ from $R_t$ $\equiv$ $\sigma_{t\overline{t}}/\sigma_{\mu^+\mu^-}$ 
{\it above} threshold.
For $Q$ $\geq$ 500 GeV the uncertainty on $R_t$ due to $m_t$ is around 0.5\%. 
The limiting precision on $R_t$ will be given by the uncertainty on
the luminosity measurement. 
If this is as good as 0.5\%~\cite{tdr} 
then $\alpha_s(M_Z)$ could be determined with an experimental precision 
approaching 1\%, which would be extremely valuable as a complementary 
precision measurement from the $t\overline{t}$ system.

\subsection{A High-luminosity Run at the $Z^0$ Resonance}

A Giga $Z^0$ sample offers two additional options for 
$\alpha_s(M_Z)$ determination via measurements of the inclusive ratios 
$\Gamma^{had}_Z/\Gamma_Z^{lept}$ and
$\Gamma^{had}_{\tau}/\Gamma_{\tau}^{lept}$. Both are indirectly 
proportional to $\alpha_s$, and hence require a very large event sample for 
a precise measurement. For example, the current LEP data sample of 16M
$Z^0$ yields an error of 2.5\% on $\alpha_s(M_Z)$ from 
$\Gamma^{had}_Z/\Gamma_Z^{lept}$. The statistical error could,
naively, be pushed to below the $\Delta\alpha_s(M_Z)$ = 0.4\% level, but systematic errors
arising from the hadronic and leptonic event
 selection will probably limit the precision
to 0.8\%~\cite{tdr}. 
This would be a very precise, reliable 
measurement. In the case of 
$\Gamma^{had}_{\tau}/\Gamma_{\tau}^{lept}$ the experimental precision
from LEP and CLEO is already at the 1\% level on $\alpha_s(M_Z)$. However,
there has been considerable debate about the size of the theoretical
uncertainties, with estimates as large as 5\% \cite{tau}. 
If this situation is clarified, and the theoretical uncertainty is small,  
$\Gamma^{had}_{\tau}/\Gamma_{\tau}^{lept}$ may offer a further
1\%-level $\alpha_s(M_Z)$ measurement. 

\section{$Q^2$ Evolution of $\alpha_s$}

In the preceeding sections we discussed the expected attainable precision
on the yardstick $\alpha_s(M_Z)$. Translation of the measurements of $\alpha_s$($Q$) 
($Q \neq M_Z$) to $\alpha_s(M_Z)$ requires the assumption that the `running' of
the coupling is determined by the QCD $\beta$ function.
However, since the logarithmic decrease of $\alpha_s$ with $Q$ is an essential
component of QCD, reflecting the underlying non-Abelian dynamics,
it is vital also to test this $Q$-dependence explicitly.
Such a test would be particularly interesting if new coloured particles were 
discovered, since deviations from QCD running would be expected at 
energies above the threshold for pair-production of the new particles.
Furthermore, extrapolation of $\alpha_s$ to very high energies of the 
order of $10^{15}$ GeV can be combined 
with corresponding extrapolations of the
dimensionless weak and electromagnetic couplings in order to
constrain the coupling-unification, or GUT, scale~\cite{gut}. 
Hence it would be desirable to measure $\alpha_s$ {\it in the same detector,
with the same technique, and by applying the same treatment to the
data} at a series of different energies $Q$, so as to maximise the
lever-arm for constraining the running.

\begin{figure}
 \includegraphics[width=140mm]{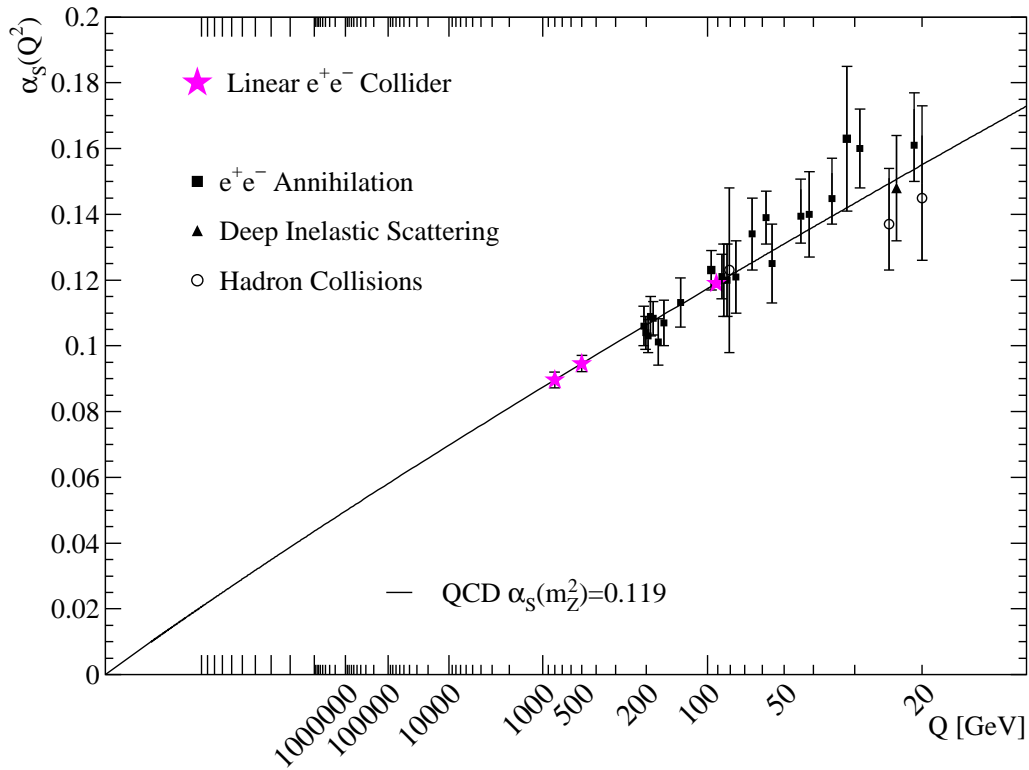}
    \caption{The evolution of $\alpha_s$ with $1/\ln Q$ \protect{\cite{otmar}}; 
             sample $Q$ values (GeV) are indicated.}
 \label{tdralphas}
 \end{figure}

Simulated measurements of $\alpha_s$($Q$) at $Q$ = 91, 500 and 800 GeV
are shown in Fig.~\ref{tdralphas}, together with
existing measurements which span the range $20\leq Q\leq 200$ GeV.
The highest-energy measurements are currently provided by LEPII. 
The point at $Q$ = 91 GeV is based on the $\Gamma_Z^{had}/\Gamma_Z^{lept}$
technique, and those at 500 and 800 GeV are based on the 
event shapes technique.  The last two include the current theoretical uncertainty,
which yields a total error on each point equivalent to $\Delta\alpha_s(M_Z)$ = 4\%.
It is clear that the LC data would add significantly to the
lever-arm in $Q$, and would allow a substantially improved extrapolation
to the GUT scale. 

\section{Further Important Topics}

Limited space allows only a brief mention
of several other important topics~\cite{previous}:

\begin{itemize}

\item
{\it Hard gluon radiation in $t\overline{t}$  events} would allow several 
tests of the strong dynamics of the top quark~\cite{arnd}:
test of the flavour-independence of strong interactions;
limits on anomalous chromo-electric and/or
chromo-magnetic dipole moments~\cite{rizzo};
determination of the running $m_t$.

\item
{\it Soft gluon radiation in $t\overline{t}$ events} 
is expected to be strongly regulated by
the large mass and width of the top quark. Precise measurements of
gluon radiation patterns in $t\overline{t}g$  events would provide 
additional constraints on the top decay width~\cite{lynne}.

\item
{\it Polarised electron (and positron) beams} can be exploited 
to test symmetries using multi-jet final states.
For polarized $e^+e^-$ annihilation to
three hadronic jets one can define ${\bf S}_e\cdot({\bf k}_1\times {\bf k}_2)$,
which correlates the electron-beam polarization vector ${\bf S}_e$
with the normal to the three-jet plane defined by
${\bf k}_1$ and ${\bf k}_2$, the momenta of the two quark jets.
If the jets are ordered by momentum (flavour)
the triple-product is CP even (odd) and T odd.
Standard Model T-odd contributions of this form are
expected~\cite{lance} to be immeasurably small, and limits 
have been set for the $b\overline{b}g$ system~\cite{sldtodd}.
At the LC these observables will provide
an additional search-ground for anomalous effects in the $t\overline{t}g$
system.

\item
{\it The difference between the particle multiplicity in heavy- ($b,c$)
and light-quark events} is predicted~\cite{doksh} to be independent of
c.m. energy. Precise measurements have been made at the $Z^0$, but 
measurements at other energies are statistically limited in precision, 
rendering a
limited test of this important prediction. High-precision
measurements at the LC would add the lever-arm for a powerful test.

\item
{\it Colour reconnection and Bose-Einstein correlations}
are important to study precisely since
they may affect the precision with which the masses of heavy particles,
such as the $W^{\pm}$ and top-quark, can be reconstructed kinematically
via their multijet decays~\cite{torbj}.

\item
{\it Hadronisation studies and renormalon physics}
can be explored via measurements of event-shape observables over a
range of $Q$ values.

\end{itemize}

\section{Summary}
There is a rich programme of QCD studies at the Linear Collider.
Precision measurements of the strong coupling and of the strong dynamics
of the $t\overline{t}$ system will complement inclusive measurements that
will be made at the LHC.

\end{document}